\documentclass[twocolumn,showpacs,preprintnumbers,amsmath,amssymb]{revtex4}
\usepackage[dvips]{graphicx}
\usepackage{dcolumn}
\begin{document}


\title
{Antiferromagnetic ordering induced by paramagnetic depairing in unconventional superconductors}

\author{Ryusuke Ikeda, Yuhki Hatakeyama, Kazushi Aoyama}

\affiliation{%
Department of Physics, Kyoto University, Kyoto 606-8502, Japan
}

\date{\today}


\begin{abstract} 
Antiferromagnetic (AFM) (or spin-density wave) quantum critical fluctuation enhanced just below $H_{c2}(0)$ have been often observed in $d$-wave superconductors with a strong Pauli paramagnetic depairing (PD) including CeCoIn$_5$. It is shown here that such a tendency of field-induced AFM ordering is a consequence of strong PD and should appear particularly in superconductors with a gap node along the AFM modulation. Two phenomena seen in CeCoIn$_5$, the anomalous vortex lattice form factor and the AFM order in the Fulde-Ferrell-Larkin-Ovchinnikov state, are explained based on this peculiar PD effect. 
\end{abstract}

\pacs{}


\maketitle


Recently, the heavy-fermion $d$-wave paired superconductor CeCoIn$_5$ with strong paramagnetic depairing (PD) has been thoroughly studied from the viewpoint of identifying its high field and low temperature (HFLT) phase near the zero temperature depairing field $H_{c2}(0)$ with a possible Fulde-Ferrell-Larkin-Ovchinnikov (FFLO) state \cite{Bianchi,AI1}. On the other hand, this material also shows an antiferromagnetic (AFM) or equivalently, a spin-density wave ordering \cite{Kenzelmann} in the HFLT phase and transport phenomena suggestive of an AFM quantum critical point (QCP) lying near $H_{c2}(0)$ \cite{Capan,Kasahara}. A similar AFM fluctuation enhanced near $H_{c2}(0)$ has also been detected in other heavy-fermion superconductors \cite{Park,Haga} and cuprates \cite{Shibauchi}, all of which seem to have strong PD and a $d$-wave pairing. A conventional wisdom on this issue will be that, in zero field, the nonvanishing superconducting (SC) energy gap suppresses AFM ordering and thus that the field-induced reduction of the gap leads to a recovery of AFM fluctuation. However, it is difficult to explain, based on this picture, why an apparent QCP is realized not above but only just below $H_{c2}(0)$ \cite{Capan,Kasahara} in those materials. Rather, the fact that the field-induced AFM ordering or QCP close to $H_{c2}(0)$ is commonly seen in superconductors with a $d$-wave pairing and strong PD suggests a common mechanism peculiar to superconductivity in finite fields and independent of electronic details of those materials such as the band structure. 

In this paper, we point out that, in nodal $d$-wave superconductors, a field-induced enhancement of PD tends to induce an AFM ordering just below $H_{c2}(0)$. Although relatively weak PD tends to be suppressed by the quasiparticle damping effect brought by the AFM fluctuation, strong PD rather favor coexistence of a $d$-wave superconductivity and an AFM order. Detailed mechanism of this field-induced enhancement of AFM fluctuation or ordering below $H_{c2}$ depends upon the relative orientation between the moment ${\bf m}$ of the expected AFM order and the applied field ${\bf H}$ : In ${\bf m} \parallel {\bf H}$ case, strong PD change the sign of the ${\rm O}(m^2 |\Delta|^2)$ term in the free energy for any pairing symmetry, just like that of its ${\rm O}(|\Delta|^4)$ term leading to the first order $H_{c2}$-transition \cite{AI1}, where $m \equiv |{\bf m}|$ and $\Delta$ are order parameters of an expected AFM phase and a spin-singlet SC one. In contrast, the field-induced AFM ordering in ${\bf m} \perp {\bf H}$, which is possibly satisfied in CeCoIn$_5$ in ${\bf H} \perp c$ \cite{Kenzelmann}, is a peculiar event to the $d$-wave paired case with the momentum (${\bf k}$) -dependent gap function $w_{\bf k}$ satisfying $w_{\bf k} = - w_{{\bf k}+{\bf Q}}$ and tends to occur irrespective of the presence of the first order $H_{c2}$-transition, where ${\bf Q}$ is the wave vector of the commensurate AFM modulation. As a consequences of this PD-induced magnetism, two striking phenomena observed in CeCoIn$_5$, an AFM order \cite{Kenzelmann} stabilized by a FFLO spatial modulation and an anomalous flux density distribution in the vortex lattice \cite{Science}, will be discussed. 

First, we start from a BCS-like electronic Hamiltonian \cite{Ueda} for a quasi two-dimensional (2D) material with $\Delta$ and ${\bf m}$ introduced as functionals. By treating $\Delta$ at the mean field level, the free energy expressing the two possible orderings is written in zero field case as 
\begin{eqnarray}
{\cal F}({\bf H} \!\!\! &=& \!\!\! 0) = \sum_{\bf r} \frac{1}{g} |\Delta({\bf r})|^2 - T \, {\rm ln} \, {\rm Tr}_{c, c^\dagger, m} 
\exp[- H_{\Delta \, m}/T], \nonumber \\
H_{\Delta \, m} &=& \sum_{\alpha,\beta=\uparrow,\downarrow} \biggl( \sum_{\bf k}  \, {\hat c}^\dagger_{{\bf k}, \alpha} \, e_{\bf k} \, \delta_{\alpha,\beta} \, {\hat c}_{{\bf k},\beta} - \sum_{\bf q} [ {\bf m}({\bf q})\cdot{\hat {\bf S}}^\dagger({\bf q}) \nonumber \\
&+& {\rm h.c.} \, ] \biggr) + 
\sum_{\bf q} \biggl(\frac{1}{U} |{\bf m}({\bf q})|^2 + \Delta({\bf q}) {\hat \Psi}^\dagger({\bf q}) \biggr), 
\end{eqnarray}
where ${\hat \Psi}({\bf q}) = - \, \sum_{\bf k} w_{\bf k} \, {\hat c}_{-{\bf k}+{\bf q}/2, \uparrow} {\hat c}_{{\bf k}+{\bf q}/2, \downarrow}$, ${\hat S}_{\nu}({\bf q})= (\sigma_\nu)_{\alpha,\beta} \, \sum_{\bf k} {\hat c}^\dagger_{{\bf k}-{\bf q}, \alpha} {\hat c}_{{\bf k}+{\bf Q}, \beta}/2$, ${\hat c}^\dagger_{{\bf k}, \alpha}$ creates a quasiparticle with spin index $\alpha$ and momentum ${\bf k}$, $\sigma_\nu$ are the Pauli matrices, and the positive parameters $g$ and $U$ are the attractive and repulsive interaction strengths leading to the SC and AFM orderings, respectively. The dispersion $e_{\bf k}$, measured from the chemical potential, satisfies $e_{\bf k} = - e_{{\bf k}+{\bf Q}} + T_c \, \delta_{\rm I}$ \cite{Chubukov}, where a small parameter $\delta_{\rm I}$ measuring an incommensurability of the AFM order was introduced. In the case with a nonzero ${\bf H}$ of our interest, the Zeeman energy $\mu_{\rm B} {\bf H}\cdot({\bf \sigma})_{\alpha,\beta}$ needs to be added to $e_{\bf k} \delta_{\alpha,\beta}$. Below, we focus on either the case, ${\bf m} \perp \, {\bf H}$ or ${\bf m} \parallel \, {\bf H}$ by choosing the spin quantization axis along ${\bf H}$. The orbital field effect will be included later. 

\begin{figure}[t]
\scalebox{1.2}[1.1]{\includegraphics{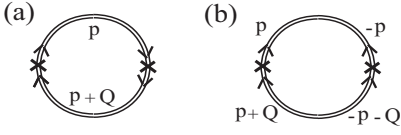}}
\caption{Diagrams describing (a) $\chi^{({\rm n})}$ and (b) $\chi^{({\rm an})}$, where the cross denotes the particle-hole vertex on the AFM fluctuation, and a double solid line in (a) and (b) denotes a normal and anomalous Green's function, respectively. }
\label{fig.1}
\end{figure}

To examine an interplay between the AFM and SC orderings below, let us consider the Gaussian AFM fluctuation term ${\cal F}_{m}$ in the free energy ${\cal F}$, ${\cal F}_m = T \sum_{\Omega} {\rm ln} \, {\rm det}[ \, U^{-1} \delta_{{\bf q}, {\bf q}'} - \chi_{{\bf q}, {\bf q}'}(\Omega) \, ]$, where 
\begin{eqnarray}
\chi_{{\bf q}, {\bf q}'}(\Omega) &=& \! \int_0^{T^{-1}} d\tau \, \langle \, T_\tau \, {\hat S}^\dagger_\nu({\bf q};\tau) \, {\hat S}_\nu({\bf q}';0) \, \rangle \, e^{i \Omega \tau}, 
\end{eqnarray}
with a fixed $\nu$, and 
${\hat S}_\nu({\bf q};\tau)$ denotes ${\hat S}_\nu({\bf q})$ at imaginary time $\tau$. 
For the moment, we focus on the Pauli limit with no orbital field effect and with uniform $\Delta$ in which $\chi_{{\bf q}, {\bf q}'}(\Omega) = [\chi^{({\rm n})}({\bf q}, \Omega) + \chi^{({\rm an})}({\bf q}, \Omega)] \delta_{{\bf q}, {\bf q}'}$, and ${\cal F}_{m} = - T \sum_{{\bf q}, \Omega} {\rm ln} X({\bf q}, \Omega)$, where $X^{-1}({\bf q}, \Omega) = U^{-1} - \chi^{({\rm n})}({\bf q}, \Omega) - \chi^{({\rm an})}({\bf q}, \Omega)$, and $\chi^{({\rm n})}$ and $\chi^{({\rm an})}$ are expressed by Fig.1 (a) and (b), respectively. The $\Delta$-dependent term, $\chi_s({\bf q}, \Omega) = \chi^{({\rm n})}({\bf q}, \Omega) - \chi^{({\rm n})}({\bf q}, \Omega)|_{\Delta=0} + \chi^{({\rm an})}({\bf q}, \Omega)$, has been studied in literatures \cite{Ueda} in $H=0$ case. The corresponding result of Fig.1 in a nonzero $H$ is given, in the Pauli limit, by 
\begin{eqnarray}
\chi^{({\rm n})}(0, \Omega) &=& - T \sum_{\bf p} \sum_{\varepsilon, \sigma = \pm 1}  \frac{d_{\varepsilon+\Omega,{\overline \sigma}}(e_{{\bf p}+{\bf Q}}) \, d_{\varepsilon,\sigma}(e_{{\bf p}})}{D_{\varepsilon+\Omega,{\overline \sigma}}({\bf p}+{\bf Q}) \, D_{\varepsilon,\sigma}({\bf p})}, \nonumber \\
\chi^{({\rm an})}(0, \Omega) &=& T \sum_{\bf p} \sum_{\varepsilon, \sigma = \pm 1} \frac{( - w_{\bf p} w_{{\bf p}+{\bf Q}}) \, |\Delta|^2}{D_{\varepsilon+\Omega,{\overline \sigma}}({\bf p}+{\bf Q}) \, D_{\varepsilon,\sigma}({\bf p})}, 
\label{nan}
\end{eqnarray}
where $\varepsilon$ ($\Omega$) is a fermion's (boson's) Matsubara frequency, $d_{\varepsilon,\sigma}(e_{\bf p}) = {\rm i}\varepsilon + \sigma \mu_{\rm B} H + e_{\bf p}$, and $D_{\varepsilon,\sigma}({\bf p}) = -d_{\varepsilon,\sigma}(e_{{\bf p}}) \, d_{\varepsilon,\sigma}(-e_{{\bf p}})+ |\Delta w_{\bf p}|^2$.  
Main features of $\chi^{({\rm n})}$ and $\chi^{({\rm an})}$ are seen in their O($|\Delta|^2$) terms. In zero field, $\chi_s(\Delta) \equiv \chi_s(0,0)$ behaves like $T^{-2}$ in $T \to 0$ limit and is negative so that the AFM ordering is suppressed by superconductivity \cite{Ueda}. 

To explain results in the case of strong PD, let us first focus on ${\bf m} \parallel {\bf H}$ case in which ${\overline \sigma}=\sigma$. For a near-perfect nesting, the O($|\Delta|^2$) terms of $\chi^{({\rm n})} - \chi^{({\rm n})}|_{\Delta=0}$ and $\chi^{({\rm an})}$, at $|{\bf q}| = \Omega = 0$, take the same form as the coefficient of the quartic (O($|\Delta|^4$)) term of the SC Ginzburg-Landau (GL) free energy and thus, change their sign upon cooling \cite{MT}. Thus, $\chi_s(\Delta)$ becomes positive for stronger PD, leading to a lower ${\cal F}_m$, i.e., an {\it enhancement} of the AFM ordering in the SC phase. As well as the corresponding PD-induced sign-change of the O($|\Delta|^4$) term which leads to the first order $H_{c2}$-transition \cite{AI1}, the PD-induced positive $\chi_s$ is also unaffected by inclusion of the orbital depairing. 

In ${\bf m} \perp {\bf H}$ where ${\overline \sigma}=-\sigma$, a different type of PD-induced AFM ordering occurs in a $d$-wave pairing case with a gap node along ${\bf Q}$ where $w_{{\bf p}+{\bf Q}} w_{\bf p} < 0$ : In this case, the O($|\Delta|^2$) term of $\chi^{({\rm n})}(0,0)$ remains negative and becomes $-N(0)|\Delta|^2/[2 (\mu_{\rm B} H)^2 ]$ in $T \to 0$ limit with no PD-induced sign change, where $N(0)$ is the density of states in the normal state. Instead, the corresponding $\chi^{({\rm an})}(0,0)$ and thus, $\chi_s$ are divergent like $N(0) [ \, |\Delta|/(\mu_{\rm B} H) \,]^2 \, |{\rm ln}[{\rm Max}(t, |\delta_{\rm I}|)]|$ in $T \to 0$ limit while keeping their positive signs, where $t=T/T_c$. This divergence is {\it unaffected} by including the orbital depairing. That is, in the $d_{x^2-y^2}$-wave case with ${\bf Q}=$ ($\pi$, $\pi$) \cite{com}, the AFM order tends to occur upon cooling even in ${\bf m} \perp {\bf H}$.  In contrast, $\chi^{({\rm an})}(0,0)$ is also negative in the $d_{xy}$-wave case satisfying $w_{\bf p} w_{{\bf p}+{\bf Q}} > 0$ so that the AFM ordering is rather suppressed with increasing $H$. 

\begin{figure}[t]
\scalebox{0.7}[0.55]{\includegraphics{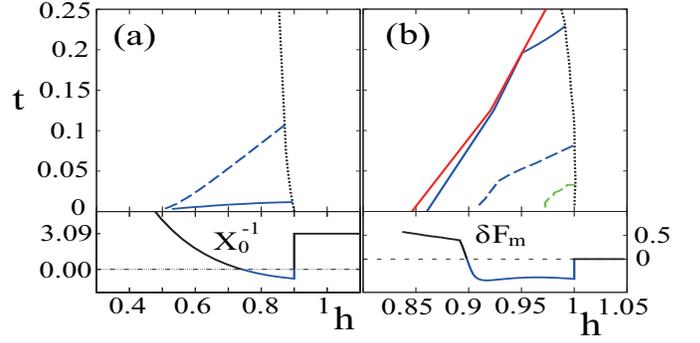}}
\caption{(Color online) Possible $\chi_s=0$ lines (dashed curves) in $t$ v.s. $h \equiv H/H_{c2}(0)$ diagram in ${\bf H} \perp c \parallel {\bf m}$ following (a) from the O($|\Delta|^2$) term of $\chi_s(\Delta)$ for $\delta_{\rm I}=0$ and (b) from the full $\chi_s(\Delta)$ in the Pauli limit for $\delta_{\rm I}=0.44$ (lower dashed (green) curve) and $0.63$ (higher dashed (blue) one), respectively. With no AFM phase in $H  > H_{c2}(T)$ (dotted curve), a dashed curve is the upper limit of a possible uniform AFM phase. In (a), the solid curve is a uniform AFM phase boundary following from the dashed curve and an assumed $X_0$ in the normal state. The lower panel of (a) is the corresponding $X_0^{-1}(h)$ at $t=0.0075$. However, the uniform AFM order parameter $m$ in the uniform $\Delta$ is transformed, in the modulated FFLO state with $\Delta(\zeta)$, to the modulated one $m(\zeta)$: In (b), the upper limit (lower solid (blue) curve) of the {\it modulated} AFM phase for $\delta_{\rm I}=0.63$ lies close to the second order FFLO transition \cite{RI07} (higher solid (red)) line. The lower panel of (b) is the normalized $\delta {\cal F}_m^{({\rm MF})}(h)$ at $t=0.075$.} 
\label{fig.2}
\end{figure}

A possible AFM phase boundary in ${\bf m} \parallel c \perp {\bf H} $ case defined as the temperature at which $X_0^{-1} \equiv [X(0,0)]^{-1}$ vanishes up to O($|\Delta|^2$) is shown in Fig.2(a) together with the corresponding $\chi_s=0$ curve, which is the upper limit of a PD-induced AFM phase to occur and shifts to {\it higher} temperature for larger $|\delta_{\rm I}|$. 
Here, the orbital depairing brought by the gauge field ${\bf A}$ has been incorporated through the quasiparticle Green's function ${\cal G}_{\varepsilon,\sigma}({\bf r}_1, {\bf r}_2)$ in real space with its semiclassical replacement \cite{Werthamer,AI1} ${\cal G}_{\varepsilon,\sigma}({\bf r}_1, {\bf r}_2) \equiv {\cal G}_{\varepsilon,\sigma}({\bf r}_1 - {\bf r}_2) \exp[i \, e (\hbar c)^{-1} \int_{{\bf r}_2}^{{\bf r}_1} {\bf A}\cdot d{\bf l} ]$. In Fig.2(a), we have used $\alpha_{\rm M}^{({\rm ab})}$ (Maki parameter in ${\bf H} \perp c$) $\equiv 7.1 \mu_{\rm B} H_{c2,\perp}^{({\rm GL})}(0)/(2 \pi T_c)=7.8$ and the anisotropy $\gamma$ (defined \cite{RI07} from $e_{\bf k}$) $= 4.5$, where $H_{c2,\perp}^{({\rm GL})}(0) = \gamma H_{c2,\parallel}^{({\rm GL})}(0) = \gamma \hbar c/(2e \xi_0^2)$ is the $H_{c2}(0)$-value in ${\bf H} \perp c$ defined near $T_c$ with the coherence length $\xi_0$. 
The AFM phase is lost in $H > H_{c2}$ due to the discontinuous $H_{c2}$-transition \cite{AI1} in $t < 0.215$. With a larger $X_0^{-1}$ in the normal state, the uniform AFM phase boundary in Fig.2(a) is pushed down to $t=0$ to reduce to an AFM-QCP. 

The limitation to the O($|\Delta|^2$) term of $\chi_s$ overestimates the AFM ordering. In fact, using the {\it full} expression of $\chi_s$, the AFM region for $\delta_{\rm I}= 0$ is limited to an invisibly narrow region in the vicinity of $H_{c2}(0)$. Nevertheless, the PD-induced AF order for nonzero $|\delta_{\rm I}|$ also follows from the {\it full} $\chi_s(\Delta)$: By substituting $|\Delta|$ obtained from the gap equation into eq.(3), the lines on which the full $\chi_s(\Delta)$ vanishes are obtained, in the Pauli limit with uniform $\Delta$, as the dashed curves in Fig.2(b). The $\chi_s > 0$ region becoming wider with {\it increasing} $|\delta_{\rm I}|$ suggests that, as in CeCoIn$_5$ \cite{Kenzelmann}, the PD-induced AFM order near $H_{c2}(0)$ should be incommensurate. Similar results also follow from the direct use of the tight-binding model \cite{sces}. 

Interestingly, the AFM ordered region is expanded further by the presence of the FFLO modulation $\Delta(\zeta) \equiv \sqrt{2} {\rm cos}(q_{\rm LO} \zeta)$, where $\zeta$ is the component parallel to ${\bf H}$ of the coordinate. To demonstrate this, the gradient expansion \cite{Werthamer} will be applied to the O($m^2$) term of the {\it mean field} expression of ${\cal F}_m$. Up to the lowest order in the gradient, its $\Delta(\zeta)$-dependent part simply becomes 
\begin{equation}
\delta {\cal F}_m^{({\rm MF})} = \int d\zeta \, m(\zeta) \biggl[ - (X_0^{-1})^{(2)} \frac{\partial^2}{\partial \zeta^2} - \chi_s(\Delta(\zeta)) \biggr] 
m(\zeta), 
\end{equation} 
where $(X_0^{-1})^{(2)} = \partial X^{-1}({\bf k},0)/\partial k^2|_{{\bf k}=0}$, and we only have to examine its sign which, for uniform $\Delta$, corresponds to that of $-\chi_s$. For simplicity, by using the $q_{\rm LO}$-data in Ref.\cite{RI07}, we find that $\delta {\cal F}_m^{({\rm MF})} < 0$ below the blue solid curve in Fig.2(b), indicating that the FFLO order can coexist with the AFM order in most temperature range. Close to the FFLO transition (red solid) curve on which $q_{\rm LO}=0$, $m(\zeta)$ is found to have the modulation $\sim {\rm sin}(q_{\rm LO}\zeta)$, so that this AFM order is {\it continuously} lost as the FFLO nodal planes goes away from the system \cite{Yanase}. Oppositely, the form of $m(\zeta)$ {\it deep} in the FFLO state may not be examined properly in terms of the present gradient expansion and will be reconsidered elsewhere. 

Next, we examine the anomalous flux density distribution in the vortex lattice of CeCoIn$_5$ in ${\bf H} \parallel c$ \cite{Science} as another phenomenon suggestive of an AFM fluctuation enhanced just below $H_{c2}$. The flux distribution is measured by the form factor $|F|$, which is the Fourier component of the longitudinal magnetization $M_c({\bf r}) \equiv (B_c({\bf r}) - H )/(4 \pi)$ at the shortest reciprocal lattice vector ${\bf K}$ and implies the slope of the flux density $B_c$ in real space. Here, $M_c({\bf r}) = \sum_{{\bf K} \neq 0} M_c({\bf K}) e^{i{\bf K}\cdot{\bf r}}$ is given by 
\begin{equation}
M_c({\bf K}) = {\rm i} c^{-1} K^{-2} [ {\bf K} \times {\bf j}({\bf K}) ]_c - \frac{\delta \, {\cal F}}{\delta B_{\rm pd}(-{\bf K})} \biggl|_{B_{\rm pd}=0}, 
\end{equation}
where ${\bf j}({\bf K})$ and $B_{\rm pd}({\bf K})$ are Fourier components of the orbital supercurrent density and a Zeeman field $B_{\rm pd}$ imposed to define $M_c$, respectively. It is straightforward to, based on the approach in Ref.\cite{AI1}, obtain $M_c$ in the weak-coupling model with no AFM fluctuation by fully taking account of the PD and the orbital depairing. In our numerical calculations, the ${\rm O}(|\Delta|^4)$ contributions to $M_c$ are also incorporated. Nevertheless, it is instructive to first focus on its O($|\Delta|^2$) terms. In $H \ll H_{c2,\parallel}^{({\rm GL})}(0)$, the contribution to $M_c$ from the spin part (last term) of eq.(5) is given by 
\begin{eqnarray}
M_c({\bf K})^{({\rm pd})}= \frac{\mu_B N(0)}{2 \pi T} |\Delta|^2({\bf K}) {\rm Im} \, \psi^{(1)}\biggl(\frac{1}{2} 
+ {\rm i} \frac{\mu_B H}{2 \pi T} \biggr)  
\end{eqnarray}
which is negative, where $\psi^{(1)}(z)$ is the first derivative of the digamma function. Thus, the PD tends to enhance the magnetic screening far from the vortex core. This is one of origins inducing a field-induced increase of $|F|$. However, such an enhancement of $|F|$ calculated beyond the O($|\Delta|^2$) terms in the weak coupling model is much weaker in the $d$-wave case than in the $s$-wave one \cite{IM1}, although it is the $d$-wave material CeCoIn$_5$ which has clearly shown such an enhanced $|F|$ \cite{Science}. This has motivated us to see how the PD-induced AFM fluctuation is reflected in $|F|$. 
For this purpose, let us first consider the ${\bf m} \perp {\bf H}$ case and start with the Pauli limit again in which $M_c$ is determined by the last term of eq.(5). By noting that this term can be obtained as the first derivative of the free energy {\it density} with respect to $\mu_{\rm B} H$, it is found that the contribution to $M_c$ from the Gaussian AFM fluctuation is positive and proportional to $\sim {\rm O}(|\Delta({\bf r})|^2) \mu_{\rm B}^2 H \sum_{\bf q} X({\bf q},0) /T^3$, implying a suppressed screening due to the AFM fluctuation, for weak PD ($\mu_{\rm B} H \ll T$), while it is negative and given by $- 2 \mu_{\rm B} T |\Delta({\bf r})|^2 (\mu_{\rm B} H)^{-3} \, |{\rm ln}[{\rm Max}(|\delta_{\rm I}|, t)]| \, \sum_\Omega \sum_{\bf q} X({\bf q}, \Omega)$, suggesting an enhancement due to the AFM fluctuation of the screening and thus, of $|F|$, for strong PD ($\mu_{\rm B} H \gg T$), respectively. 

In Fig.3, we show $h$ v.s. $|F|^2$ curves obtained by incorporating the orbital depairing for both cases with and with no contributions to eq.(5) from the AFM fluctuation through ${\cal F}_m$, where $|F|$ was normalized by that in the GL region near $T_c$. Based on the result in Fig.2(a), the familiar phenomenological form of $X({\bf q},\Omega)$, $N(0) X({\bf q},\Omega) = 1/[ m_{\rm N}(t,h) + (|{\bf q} \times {\hat z}| \, \xi^{({\rm N})}_0)^2 + \xi^{({\rm N})}_0 |\Omega|/|v|]$ was assumed in calculating the ${\cal F}_m$-contribution to eq.(5), where $m_{\rm N}(t,h) = t+1-h/h_{\rm QCP}$, and $\xi^{({\rm N})}_0 = 0.6 \xi_0$, and $v^2$ is the squared Fermi velocity averaged over the Fermi surface. 
The remarkable peak just below $H_{c2}$ of the red solid curve is a reflection of the afore-mentioned growth of $\chi^{({\rm an})}$ in ${\bf m} \perp {\bf H}$. A similar $|F|$-enhancement also occurs in ${\bf m} \parallel {\bf H}$ (dashed) curves with increasing $H$, reflecting the afore-mentioned sign change of $\chi_s$ due to large PD. To the best of our knowledge, the favorable direction of the moment ${\bf m}$ in CeCoIn$_5$ in ${\bf H} \parallel c$ is unclear at present. If, in CeCoIn$_5$, the ${\bf m} \perp {\bf H}$ components are dominant in the AFM fluctuation even in ${\bf H} \parallel c$, the $|F(t,h)|$-data \cite{Science} growing with increasing field {\it and} on further cooling is a reflection of the $d_{x^2-y^2}$-wave pairing \cite{HI} accompanied by a strong AFM fluctuation with ${\bf Q} \parallel (\pi$, $\pi)$. With the {\it same} ${\bf Q}$, the corresponding $|F(t,h)|$ in the $s$-wave case would become much lower than the dotted ones, contrary to the trend in the weak-coupling results \cite{IM1}. The curves in Fig.3 have been obtained by neglecting the narrow FFLO region in CeCoIn$_5$ in ${\bf H} \parallel c$. Effects of the FFLO modulation on $M_c$ will be reported elsewhere. 

\begin{figure}[t]
\scalebox{0.52}[0.44]{\includegraphics{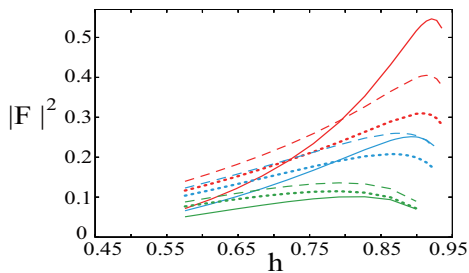}}
\caption{(Color online) Normalized form factor $|F(h)|^2$ curves in ${\bf H} \parallel c$  obtained in the weak-coupling BCS model with no AFM fluctuation (dotted curves), in the case with the AFM contribution with ${\bf m} \parallel {\bf H}$ (dashed ones), and in the corresponding ${\bf m} \perp {\bf H}$ case (solid ones) at $t=0.02$ (highest (red) curves), $0.04$ (middle(blue)), and $0.08$ (lowest (green)). Here, $h_{\rm QCP}=0.942$, $\alpha_{\rm M}^{(c)} = \alpha_{\rm M}^{(ab)}/\gamma = 6.5$, and the typical quasiparticle damping rate $\pi^2 T/[ 4 k_F \xi^{({\rm N})}_0 (m_{\rm N}(t,h))^{1/2} ]$ due to 2D AFM fluctuation was used.}
\label{fig.3}
\end{figure}

It has been argued in previous studies \cite{Bianchi,AI1,HI} that the anomalous SC properties in CeCoIn$_5$ in high fields are consequences of strong PD. The present results indicate that the AFM fluctuation and order indicated in more recent measurements \cite{Kenzelmann,Science} on this material just below $H_{c2}(0)$ also stem from PD, and thus that its HFLT phase and the behaviors suggestive of an AFM-QCP near $H_{c2}(0)$ commonly seen in $d$-wave superconductors with strong PD \cite{Capan,Kasahara,Park,Haga,Shibauchi}, including CeCoIn$_5$, should be understood on the same footing. For instance, the result in Fig.2 that the field-induced AFM ordering is {\it discontinuously} bounded by the first order $H_{c2}$-transition naturally explains why, in spite of the AFM ordering just below $H_{c2}(0)$ \cite{Kenzelmann}, the transport properties in CeCoIn$_5$ in $H > H_{c2}$ have suggested a QCP notably below $H_{c2}(0)$ \cite{Capan}. We have also shown that, in $d_{x^2-y^2}$-wave superconductors, the AFM order is significantly enhanced by the periodic potential due to the FFLO modulation parallel to ${\bf H}$ irrespective of the size of wave number $q_{\rm LO}$. In contrast, other works have {\it assumed} the presence of an {\it attractive} $\pi$-triplet pairing channel as the only factor inducing the AFM ordering below $H_{c2}$ \cite{Kenzelmann,Aperis}. However, the field-induced AFM transition in Ref.\cite{Aperis} is of first order in contrast to the observation \cite{Bianchi} in CeCoIn$_5$. The present work has shown that an incommensurate AFM ordering just below $H_{c2}(0)$ develops without assuming such an occurrence of other pairing state and due only to strong PD effect. In fact, the strange impurity effects, arising even from a nonmagnetic doping, on the HFLT phase of CeCoIn$_5$ \cite{Tokiwa} cannot be explained unless the HFLT phase has a spatial modulation of the SC order parameter unrelated to the AFM order \cite{RI10}. 

In conclusion, an AFM ordering or fluctuation enhanced close to $H_{c2}(0)$, often seen in unconventional superconductors, is a direct consequence of strong paramagnetic depairing and also of their $d$-wave pairing symmetry with a gap node parallel to the AFM modulation. The AFM ordering enhanced due to the FFLO modulation suggests that the HFLT phase in CeCoIn$_5$ is a coexistent state of the AFM and FFLO orders. Discussing the AFM-QCP issues in systems \cite{Park,Shibauchi} with a {\it continuous} behavior around $H_{c2}$ based on the present theory is straightforward and will be performed elsewhere. 

This work was supported by Grant-in-Aid for Scientific Research [No. 20102008 and 21540360] from MEXT, Japan.

\end{document}